\documentclass[a4paper]{jpconf}

\usepackage{amsmath}

\newcommand{\p}{\partial}

\begin{document}

\title{Hyperbolicity of Hamiltonian formulations in General Relativity}


\author{Ronny Richter}
\address{Mathematisches Institut, Auf der Morgenstelle 10, 72076 T\"ubingen, Germany}

\author{David Hilditch}
\address{Theoretical Physics Institute, University of Jena, 07743 Jena, Germany}


\begin{abstract}
Well-posedness of the initial (boundary) value problem is an essential
property, both of meaningful physical models and of numerical applications.
To prove well-posedness of wave-type equations their level of hyperbolicity 
is an essential ingredient. We develop helpful tools and classify a
large class of Hamiltonian versions of Einstein's equations with live gauge
conditions with respect to their hyperbolicity. Finally we find a symmetric
hyperbolic Hamiltonian formulation that allows for gauge conditions which are
similar to the puncture gauge.
\end{abstract}

\date{\today}

\section{Motivation}

Hamiltonian formulations play a crucial role in many areas of theoretical physics.
Their key properties are the exact conservation of an energy and the symplecticity
of the time evolution map. Certain properties of those systems can be obtained
in numerical evolutions as well, using either symplectic \cite{HaLW06}
or energy-preserving integrators \cite{Hairer2010}.

Hamiltonian formulations for General Relativity (GR) were first considered in the
1950's \cite{Pirani1952,Dirac-1958}. Today the ADM formulation~\cite{ADM1962} is 
the most popular one. Yet, in numerical relativity, where one is interested in an 
initial value problem (IVP), Hamiltonian formulations are rarely used as the basis 
of evolution schemes. One can identify a reason for this in the mathematical 
properties of the ADM system and the numerical schemes used.

The ADM system is composed of six evolution equations and four constraints.
Furthermore it possesses a gauge symmetry. It can be shown that the constraint
evolution system is closed, i.e. if the constraints are satisfied in the initial
hypersurface then they are satisfied for all times.

In numerical applications one usually prefers to solve for the constraints only
in the initial data, because the alternative of solving the constraints in every
timestep is computationally expensive. However, the IVP
for the dynamical subsystem of the ADM equations is ill-posed.
It is therefore not possible to build numerical schemes based on the dynamical 
ADM equations.

A widely used method to avoid this problem and to deal with well-posed IVPs is to
fix the gauge and to modify the ADM equations appropriately.
The dynamical ADM equations are supplemented by evolution
equations for the lapse and shift functions and multiples of the constraints are
added to the dynamical ADM equations, such that the IVP for the full system
is well-posed (the solution depends continuously on the given initial data).
Here we are interested in the question how this technique can be
applied while keeping the Hamiltonian structure.

\section{Hyperbolicity}

In order to construct systems with well-posed IVPs one relies on notions of 
hyperbolicity. It can be shown that strongly (symmetric) hyperbolic
systems possess well-posed initial (boundary) value problems (of course one needs
appropriate boundary conditions in the case of IBVPs)
\cite{GKO:1995,Nagy:2004td}. The advantage of
this approach is that the proof of hyperbolicity involves essentially linear
algebra. Strong and symmetric hyperbolicity are defined as follows \cite{Gundlach:2005ta}.

A first order in time, second order in space system of the form
\begin{subequations}
\label{eq:FOITSOIS_def}
\begin{align}
\p_t v &= A_1^i \partial_i v + A_1 v +A_2 w + a,\\
\p_t w &= B_1^{ij}\partial_i\partial_jv + B_1^i \partial_iv+B_1 v 
+B_2^i\partial_iw+B_2 w + b.
\end{align}
\end{subequations}
is called strongly hyperbolic if the principal symbol
\begin{align}
\label{eq:Psymbol}
P^s=\left(
\begin{array}{cc}
A_1^i s_i & A_2\\
B_1^{ij}s_i s_j & B_2^i s_i
\end{array}
\right),
\end{align}
has a complete set of eigenvectors (with real eigenvalues) that depend continuously
on the spatial  vector $s_i$.

The system \eqref{eq:FOITSOIS_def} is called symmetric hyperbolic if there exists 
a conserved positive definite energy, $E=\int\epsilon dx$ \cite{Gundlach:2005ta}
of the form
\begin{align}
\epsilon&=u_i^\dag H^{ij}(v) u_j
=
\left(
\begin{array}{c}
\p_i v\\
w
\end{array}
\right)^\dag
\left(
\begin{array}{cc}
H_{11}^{ij} & H_{12}^i\\
H_{12}^{j\,\dag} & H_{22}
\end{array}
\right)
\left(
\begin{array}{c}
\p_j v\\
w
\end{array}
\right).
\end{align}
We denote $H^{ij}$ a \emph{symmetrizer}. Furthermore if there
is a matrix $\bar H^{ij}$ such that $u_i^\dag\bar H^{ij} u_j$ is a conserved
quantity, but not necessarily positive definite then we call $\bar H^{ij}$
a \emph{candidate symmetrizer}.

\section{Hamiltonian formulations}
\label{sec:Hamiltonians}

A Hamiltonian formulation is given by the specification of a Hamiltonian 
${\cal H}(q,p)$ for the system, constructed from \emph{canonical positions} 
and \emph{momenta} $(q;p)$. Here the Hamiltonian can be expressed as the 
integral over space of a Hamiltonian density
\begin{align}
{\mathcal H}&=\int_{\Sigma_t} d^3x\,{\mathcal H}_D(q,\p_i q, p,\p_i p),
\end{align}
where the Hamiltonian density is a local function of the canonical
variables and their spatial derivatives.
One obtains the following \emph{canonical equations of motion}
\begin{subequations}
\label{eq:canon_eqn}
\begin{align}
\p_tq&= \frac{\p{\mathcal H}_D}{\p p}-\p_i
\frac{\p{\mathcal H}_D}{\p(\p_i p)},\\
\p_tp&=-\frac{\p{\mathcal H}_D}{\p q}+\p_i
\frac{\p{\mathcal H}_D}{\p(\p_i q)}.
\end{align}
\end{subequations}
A system that can be written in the form~\eqref{eq:canon_eqn}
has~\emph{Hamiltonian structure}.

In order to keep the Hamiltonian structure while modifying
the evolution system a good strategy is to modify the Hamiltonian
density appropriately and derive the equations of motion afterwards.
The key ideas in this approach were presented in \cite{Brown:2008cca}.

One starts from the ADM Hamiltonian density
\begin{align}
{\cal H}_{\textrm{ADM}}=
\int(-{\alpha} C+2\beta^iC_i)\sqrt{\gamma}
{\textrm d}^3x,\label{eqn:H_ADM},
\end{align}
where $\alpha$ is the lapse function, $\beta^i$ is the shift,
$\gamma$ is the determinant of the 3-metric and $C$, $C_i$ are
scalar and vector constraints respectively. The constraints are
\begin{align}
C &= \pi^{ij}\pi_{ij}-\frac{1}{2}\pi^i{}_i\pi^j{}_j- h R\,,&
C_i &= - 2 h_{ij} D_k\pi^{jk},
\end{align}
with the 3-metric $\gamma_{ij}$ and its canonically conjugate
momentum $\pi^{ij}$. The dynamical degrees of freedom are
$(\gamma_{ij},\pi^{ij})$.

Then in order to provide evolution equations for lapse and shift one needs
to include them in the set of dynamical degrees of freedom. This also means
that they need conjugate variables, which we denote $\sigma$ and $\rho_i$
respectively. One finds that these momenta vanish for solutions of GR,
i.e. $\sigma=0$ and $\rho_i=0$.

After modifying the phase space we must also work with a different
Hamiltonian. One adds to the ADM Hamiltonian terms that allow
for appropriate gauge conditions, but do not change the physics. We use
\begin{align}
\label{eq:general_Hamiltonian}
\mathcal H = \mathcal H_{\mathrm{ADM}} + \mathcal H_{\mathrm GHG}
+ \int d^3x \left(\Lambda\sigma +\Omega^i \rho_i\right),
\end{align}
where $\mathcal H_{\mathrm{ADM}} + \mathcal H_{\mathrm GHG}$ is the
Hamiltonian for Brown's generalized harmonic formulation \cite{Brown:2008cca}.
The terms that appear in the principal part are
\begin{align}
\nonumber
&\mathcal H_{\mathrm GHG} =
\beta^{i}\sigma D_{i}\alpha
+ \beta^{i}\rho_j\p_i\beta^j
+ \alpha^2\rho_i\Gamma^i_{jk}\gamma^{jk} - \alpha\rho^{i}D_{i}\alpha \\
&\qquad
+ \frac1{8\sqrt{\gamma}}
\big(-4 \alpha^2 \gamma_{ij}\pi^{ij}\sigma
 + \alpha^3 \sigma^2
 - 4 \alpha^3 \rho_{i} \rho_{j}\gamma^{ij}\big).
\end{align}
The terms in $\Lambda$ and $\Omega^i$ that affect the principal part of
the equations of motion are linear in the canonical momenta
and in the first spatial derivatives of the positions. With the
restrictions described in \cite{HR2010} we come to the following form
{
\allowdisplaybreaks
\begin{align}
\Lambda &=
-C_1 \alpha^2 \gamma^{-1/2} \gamma_{ij} \pi^{ij} +
C_4 \alpha^3 \gamma^{-1/2} \sigma+
C_7\left(\alpha D_i\beta^i - \frac12\alpha \gamma^{jk} \beta^i \p_i\gamma_{jk}\right)
\end{align}
and
\begin{align}
\Omega^i &=
C_2 \alpha^2 \Gamma^i_{jk} \gamma^{jk} +
C_3 \alpha^2 \Gamma^k_{kj} \gamma^{ji} -
C_5 \alpha \gamma^{ij} D_j\alpha
- C_6 \gamma^{-1/2} \alpha^3 \rho_j \gamma^{ij}.
\end{align}
}
With this choice the principal symbol depends on the shift in a trivial way,
which simplifies the hyperbolicity analysis.

The consequence of using this approach is that one cannot choose equations
of motion for lapse and shift and constraint additions independently. In fact,
if the gauge conditions are fixed then constraint additions are fixed
as well.

\section{Crucial Techniques in the hyperbolicity analysis}

Given the Hamiltonian \eqref{eq:general_Hamiltonian} with parameters $C_i$
it is straightforward to compute the corresponding equations of motion
and their principal symbol \eqref{eq:Psymbol}.

As we describe in \cite{HR2010}, since the principal symbol depends only trivially
on the shift, it has essentially the following structure
\begin{align}
P^s=
\left(
\begin{array}{cc}
0 & X\\
Y & 0
\end{array}
\right).
\end{align}
This matrix decomposes further into scalar, vector and trace-free tensor blocks
which simplifies the analysis further. The most complicated part in the proof is
then the scalar block. The corresponding sub blocks of $X$ and $Y$ both have upper
block triangular form:
\begin{align}
X,Y=
\left(
\begin{array}{cc}
A_{X,Y} & B_{X,Y}\\
0 & C_{X,Y}
\end{array}
\right),
\end{align}
with $2\times 2$ matrices $A$, $B$ and $C$. This structure can be used to reduce
the analysis completely to $2\times 2$ matrices. Finally one can identify all
three families of strongly hyperbolic formulations \cite{HR2010}.

Concerning symmetric hyperbolicity we consider only the strongly hyperbolic 
formulations, because symmetric implies strong hyperbolicity. The first step
is to construct candidate symmetrizers. This can be achieved by making an ansatz 
for the candidate and solving linear equations for the parameters in that ansatz.

Having a set of candidate symmetrizers one can ask for positive definite
matrices in that set. This is the most complicated step in the analysis
and we are not able to completely classify the formulations with respect to
their symmetric hyperbolicity.

We can however identify strongly hyperbolic formulations that are not symmetric
hyperbolic and have also found 2-parameter families of symmetric hyperbolic 
formulations.

Thus far we have treated the formulation parameters $C_i$ as constants. But one
can view them as functions of $(\gamma_{ij},\alpha,\beta^i)$ as well, which
effects neither the physics nor the level of hyperbolicity, but does
change the equations of motion. One can use this freedom to derive a formulation
that includes a gauge condition very similar to the popular puncture gauge. The 
equations of motion for lapse and shift are
\begin{align}
\p_t\alpha &= \beta^i\p_i\alpha-\mu_L\alpha^2 K 
+\frac1{4\sqrt{\gamma}}\mu_L\alpha^3\sigma,\\
\nonumber
\p_t\beta^i &= \beta^j\p_j  \beta^i+
\mu_S\gamma^{1/3}\Gamma^i_{jk}\gamma^{jk}
+ \frac13(\mu_S\gamma^{1/3}-\alpha^2)\Gamma^k_{kj}\gamma^{ij}
- \mu_S\gamma^{-1/6}\alpha \rho^i - \alpha D^i\alpha.
\end{align}
which is (near the puncture ($\alpha\rightarrow 0$)
and if the constraints
$\sigma=0$ and $\rho_i=0$ are satisfied)
very close to the puncture gauge condition
\begin{align}
\p_t\alpha&=\beta^i\p_i\alpha-\mu_L\alpha^2K,
&
\p_t\beta^i&=\beta^j\p_j\beta^i+\mu_S\tilde{\Gamma}^i-\eta\beta^i,
\label{eqn:Gamma_driver}
\end{align}
where
\begin{align}
\tilde{\Gamma}^i&=\gamma^{1/3}\Gamma^{i}{}_{jk}\gamma^{jk}
+\frac{1}{3}\gamma^{1/3}\gamma^{ji}\Gamma^k{}_{kj}.
\end{align}

\ack

This work was supported by DFG grant
SFB/Transregio~7 ``Gravitational Wave Astronomy''.


\end{document}